\documentclass[10pt,preprint]{aastex}

  \newcommand{\alfven}{Alfv\'{e}n }

  \newcommand{\EE}{\hat{E}}
  \newcommand{\LL}{\hat{L}}
  \newcommand{\ee}{\hat{e}} 
   
  \newcommand{\Bp}{\mathcal{B}_{p}}   
  \newcommand{\Bf}{\mathcal{B}_{\phi}}

\slugcomment{(~accepted to ApJL~ :~2010.March.30~) }

\shorttitle{Black Hole Aurora}
\shortauthors{M. Takahashi \& R. Takahashi}

\begin{document}


\title{ Black Hole Aurora powered by a Rotating Black Hole }


\author{Masaaki Takahashi}
\affil{%
 Department of Physics and Astronomy, Aichi University of Education,  
 Kariya, Aichi 448-8542, Japan 
}%
\email{takahasi@phyas.aichi-edu.ac.jp} 
\and
\author{Rohta Takahashi}
\affil{%
 Cosmic Radiation Laboratory, the Institute of Physical and Chemical 
 Research, 2-1 Hirosawa, Wako, Saitama 351-0198, Japan
}%




\begin{abstract}
 We present a model for high-energy emission sources generated by a 
 standing magnetohydrodynamical (MHD) shock in a black hole
 magnetosphere.  The black hole magnetosphere would be constructed
 around a black hole with an accretion disk, where a global magnetic
 field could be originated by currents in the accretion disk and its
 corona.  Such a black hole magnetosphere may be considered as a model
 for the central engine of active galactic nuclei, some compact X-ray
 sources and gamma-ray bursts. 
 The energy sources of the emission  from the magnetosphere are the 
 gravitational and electromagnetic energies of magnetized accreting
 matters and the rotational energy of a rotating black hole.  When the
 MHD shock generates in MHD accretion flows onto the black hole,  
 the plasma's kinetic energy and hole's rotational energy can convert 
 to radiative energy. In this letter, we demonstrate the huge energy
 output at the shock front by showing {\it negative energy}\/ postshock 
 accreting MHD flows for a rapidly rotating black hole.  This means that
 the extracted energy from the black hole can convert to the radiative
 energy    
 at the MHD shock front. When axisymmetric shock front is formed, we
 expect a ring-shaped region with very hot plasma near the black hole;
 the look would be like an ``aurora''. 
 The high energy radiation generated from there would carry to us the
 information for the curved spacetime due to the strong gravity.     
\end{abstract}


\section{Introduction} \label{sec:intro}

%
 A number of observations provides evidence for super-massive black
 holes in galactic centers or stellar-mass black hole in galaxies
\citep[e.g.][]{Miyoshi+95,Ghez+98,Ghez+05,Genzel+00,Schodel+03,
Bender+05,Eisenhauer+05}.  
 Several observations show high-energy phenomena around black hole
 candidates where some amount of energy in accretion flows is released. 
 In this paper, to understand the high-energy phenomena near the black
 hole and the nature of the curved spacetime due to the hole's strong
 gravity, we consider ingoing magnetohydrodynamic (MHD) flows in the
 magnetosphere of a black hole, which is considered as the transit
 region from an accretion disk to the event horizon. The global magnetic
 field in the magnetosphere could be generated by currents streaming in
 the disk and its corona around the black hole.

\begin{figure}[t]
 \begin{center}
   \includegraphics[scale=0.6,angle=0]{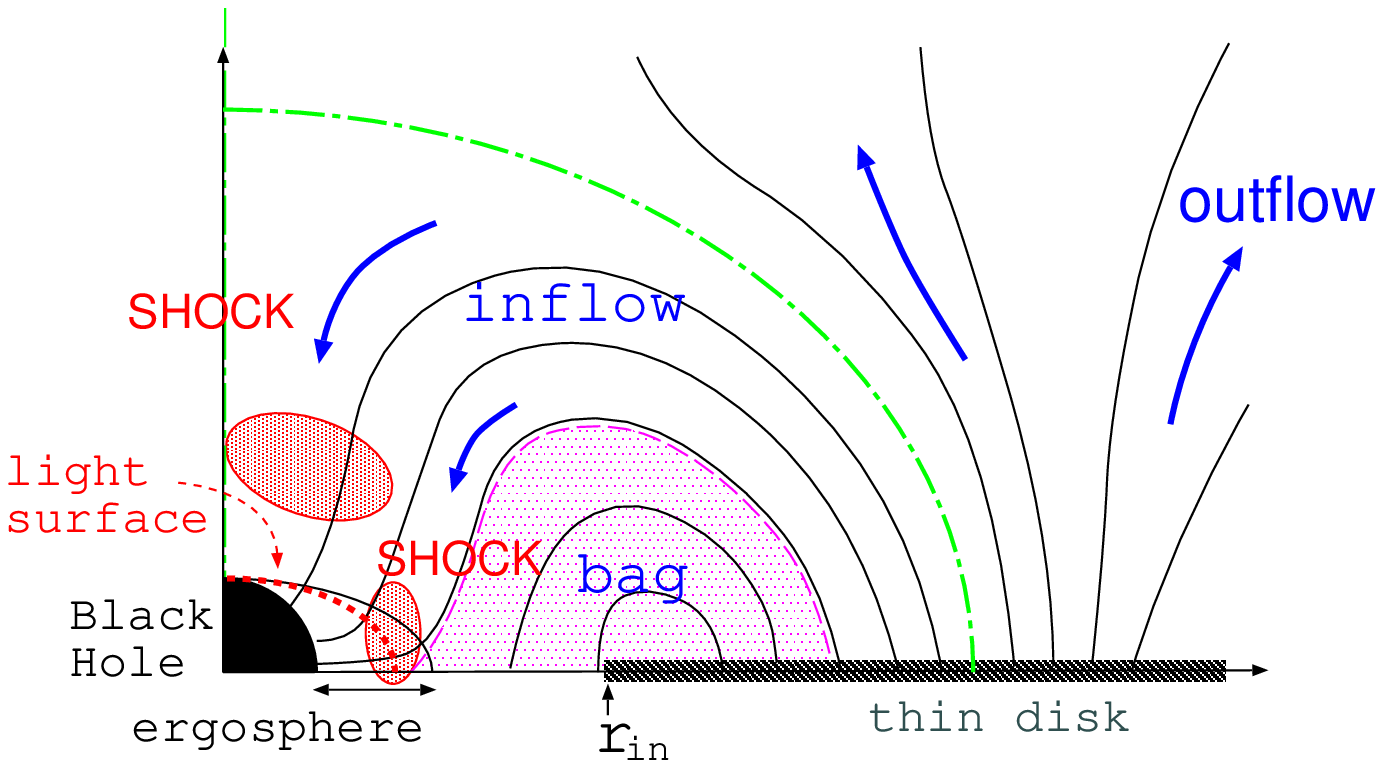}
   \includegraphics[scale=0.3,angle=0]{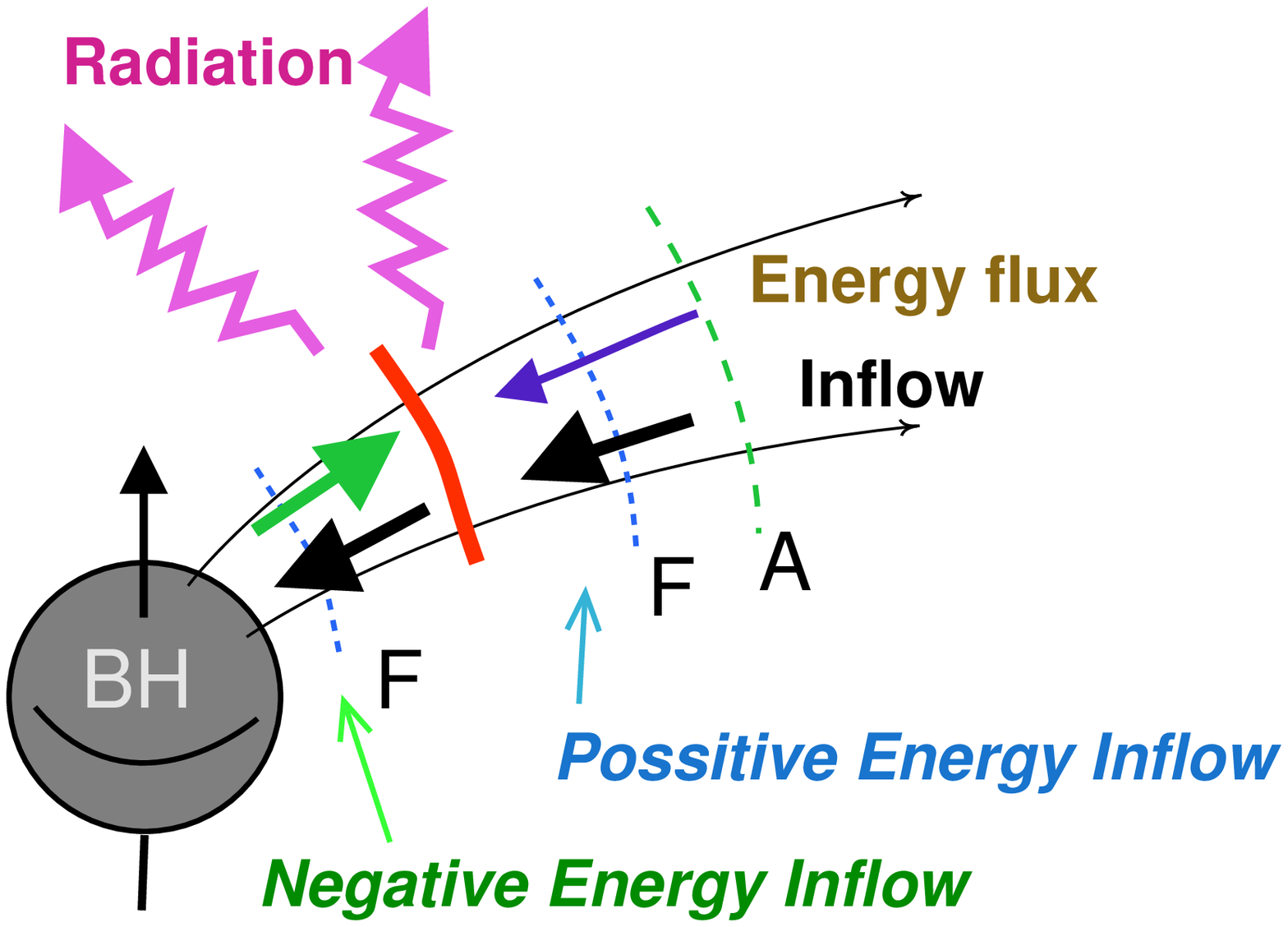}
 \end{center}
 \caption{
   (a) A model of the black hole magnetosphere. Ingoing MHD flows 
   stream along the disk--black hole connecting magnetic field lines.
   Because of the loop-shaped magnetic field configuration,
   off-equatorial shock would be expected.  
   (b) MHD shock in black hole accretion.  After passing through 
   {\it first}\/ fast magnetosonic point, the MHD inflow  can make the
   MHD shock. The postshocked sub-fast magnetosonic flows must pass
   through the {\it second}\/ fast magnetosonic point again before
   reaching the event horizon.  
 }
 \label{fig:BH-Mag}
\end{figure}

%
 The structure of a steady and axisymmetric black hole magnetosphere 
 with a thin disk has been discussed by \citet{TT01} as a vacuum
 magnetosphere, \citet{Uzdensky05} as a force-free magnetosphere, 
 and \cite{Camenzind87,NTT91} as a MHD magnetosphere. 
 These models show the disk--black hole connecting magnetic field lines,
 as shown in Figure~\ref{fig:BH-Mag}a. The loop-like shaped magnetic
 field is generated around the disk's inner edge region. Some magnetic
 field lines connect to the event horizon. Due to the plasma inertia
 effect, the magnetic field lines are forced to direct toward the event
 horizon, and then almost radial flows would be reasonable at least near
 the event horizon.   
 Such a configuration of the loop-like magnetic field lines indicates
 the interaction between the black hole and the disk surface (or corona) 
 located at several times the inner-edge radius of the thin disk.  
 Along this loop-like magnetic field lines, the MHD fluid ejected from
 the disk surface streams inward due to the dominant gravitational force
 by the black hole.  

%
 Although in many previous works equatorial accretion flows have been 
 discussed, by considering the black hole magnetosphere, we expect
 naturally off-equatorial inflows streaming along the loop-like magnetic
 field lines. So, we also expect the shock formation near the polar
 region of the black hole \cite[see][for adiabatic shocks]{FTT07}. 
 Of course, the shock formation by equatorial accretion flows would be
 possible within the required physical conditions.  However, the
 emission from the off-equatorial MHD shock would include the
 information for the ``magnetosphere'' (i.e., the configuration of the  
 magnetic fields) around a black hole, and would enable us to find the
 information for the black hole spacetime related to the radiation.

%
 We are interested in the energy conversion from the accreting plasma's
 energy, which is composed of the kinetic part and magnetic part, to the
 radiative energy at the inner-most region of the black hole
 magnetosphere.  This is because such energy conversion by the shock 
 formation would be related to the generation of huge radiative energy
 in the hole's deep gravitational potential. Furthermore, in addition to
 the plasma's kinetic energy, the rotational energy of the black hole
 may also be converted to the radiative energy (see
 Fig.~\ref{fig:BH-Mag}b). 
 For a fast MHD shock on accretion onto a black hole, about 10\% of the 
 upstream flow's energy (including the rest mass energy of a particle)
 can convert to the thermal energy of the downstream flow
 \citep{TGFRT06}. This means that the plasma temperature becomes so high
 across the shock front. 
 Such a hot plasma region can be considered as a source of high energy 
 radiation, which gives us both the information for the strong
 gravitational field and the state of magnetized plasma around the black
 hole. Of course, some part of the radiation emitted from the hot plasma
 will fall into the black hole because of the strong gravitational lens
 effects. 
\footnote{
 The escape rate of photons from the ergosphere is almost 10\% and this 
 rate becomes larger for a larger spin value \cite[see][]{RT-MT10}.  
}  
 Nevertheless we can expect that huge radiative energy can be released
 at the very hot shocked plasma region, and considerable radiation flux
 will be obtained for us.

 In this paper, we discuss energy release by shocked MHD accretion onto 
 a rapidly rotating black hole. 
 In \S 2, we introduce trans-fast magnetosonic accretion flows and apply
 the shock condition to the flows. Here, we adopt the method by
 \cite{TT08} for solving trans-fast magnetosonic flow solutions. Then,
 we can easily obtain the solutions without critical condition analysis
 at the magnetosonic points.    
 In \S 3, we show {\it negative}\/ energy postshock MHD inflows that
 means the energy extraction from a rotating black hole  
 and discuss the released energy at the shocked region.  
 In \S 4, we summarize our MHD shock models.

\section{ MHD Accretion with MHD Shock }

 We consider MHD flows in a stationary and axisymmetric magnetosphere
 around a rotating black hole.  The background metric $g_{\mu\nu}$ is
 written by the Boyer-Lindquist coordinate with $c=G=1$.  The basic
 equations for MHD flows are the equation of the particle number
 conservation, the equation of motion and Maxwell equations.  We also
 assume the ideal MHD condition and the polytropic relation for the
 plasma flows.  There are five field-aligned flow parameters on the
 flows; that is, the total energy $E$, the total angular momentum $L$,
 the angular velocity of the magnetic field line $\Omega_F$ and the
 number flux per flux-tube $\eta$ and the entropy $S$ 
 \citep[see][for the definitions of these flow
 parameters]{Camenzind86a,Camenzind86b,TNTT90}.    
 The accreting flow onto a black hole must pass through the slow
 magnetosonic point, the \alfven point, and the fast magnetosonic point,
 in this order.  The critical conditions at the \alfven point and the 
 magnetosonic points restrict the acceptable ranges of these parameters      
 \citep[see][for the details]{Takahashi02}. 
 The poloidal magnetic field $B_p$ seen by a lab-frame observer is
 defined by $B_p^2 \equiv -[ g^{rr}(F_{r\phi})^2 +
 g^{\theta\theta}(F_{\theta\phi})^2 ]/\rho_w^2$, and the toroidal
 magnetic field is defined by $B_\phi = (\Delta/\Sigma)F_{r\theta}$,
 where $\Delta=r^2-2mr+a^2$, $\Sigma=r^2+a^2\cos^2\theta$, 
 $\rho_w^2=\Delta\sin^2\theta$, and $m$ and $a$ are the hole's mass and
 spin, respectively. 
 The term $F_{\mu\nu}$ is the electromagnetic tensor.  
 In the following, we assume the cold inflows $(S=0)$ except for the
 shock front. Although, at the shock front, the plasma temperature can
 become so high, it falls off immediately by the radiation.

 The relativistic Bernoulli equation for cold MHD flows, which
 determines the poloidal velocity (or the \alfven Mach number) along a
 magnetic tube, can be written as \citep{TT08}
 \begin{equation}
     \ee^2 - \alpha - M^4 (\alpha \Bp^2 + \Bf^2) = 0 \ , 
                                                    \label{eq:pol_eq}
 \end{equation}
 where 
 $\ee\equiv \EE -\Omega_F \LL$ with 
 ${\hat E}\equiv E/\mu_{c}$ and ${\hat L}\equiv L/\mu_{c}$. 
 The enthalpy for cold flows is denoted by $\mu_{c}=m_{\rm part}$,  
 where $m_{\rm part}$ is the particle's mass.    
 The terms $\Bp \equiv B_p/(4\pi\mu_{c}\eta) $ and
 $\Bf \equiv B_\phi/(4\pi \mu_{c}\eta \rho_w) $ are introduced to
 non-dimensionalize, and the latter is given in terms of the
 relativistic \alfven Mach number and the flow parameters;   
\begin{equation}
  \Bf = \frac{G_\phi{\hat E} + G_t{\hat L}}{\rho_w(M^2 - \alpha)} \ , 
                                                    \label{eq:Bf}
\end{equation}
 where 
 $G_t \equiv g_{tt} + g_{t\phi}\Omega_F$, 
 $G_\phi \equiv g_{t\phi} + g_{\phi\phi}\Omega_F$ and 
 $\alpha \equiv G_t + G_\phi\Omega_F$.  
 The relativistic \alfven Mach number $M$ is defined by
 $ M^2 \equiv (u_p^2/u_{\rm AW}^2)\alpha = u_p^2/\Bp^2 $,    
 where $u_p$ is the poloidal velocity, the \alfven velocity 
 $u_{\rm AW}$ is defined as 
 $ u_{\rm AW}^2 \equiv \alpha B_p^2/(4\pi \mu_{c} n) $, and 
 $n$ is the number density of the plasma  
 \citep[see also][]{Camenzind86a,Camenzind86b,TNTT90}.      

 Differentiating eqs.~(\ref{eq:pol_eq}) and ~(\ref{eq:Bf}) 
 along a magnetic field line,
 we obtain the velocity gradient at any point, and see the critical
 point in flow solutions.  That is, the flow solution becomes singular
 at the magnetosonic points and the \alfven points without the
 regularity conditions. To obtain a physical solution that is smooth
 everywhere, the complicated parameter search for satisfying regularity
 conditions is required \citep{Takahashi02}. 
 To avoid the heavy task, we will introduce the ratio of the poloidal
 and toroidal components of the magnetic field as \citep{TT08} 
 \begin{equation}
             \beta \equiv \Bf/\Bp,     \label{eq:beta}
 \end{equation}
 to specify a magnetic flux-tube for the purpose of solving the
 relativistic Bernoulli equation~(\ref{eq:pol_eq}).  That is, the
 poloidal magnetic field is related to the toroidal magnetic field
 through the function $\beta(r,\theta)$, which is assumed to be regular
 at magnetosonic points, where we do not assume the concrete function
 for the poloidal magnetic field. We eliminate $\Bp^2$ in
 eq.~(\ref{eq:pol_eq}) by eq.~(\ref{eq:beta}), and also eliminate 
 $\Bf^2$ by eq.~(\ref{eq:Bf}).  Then, eq.~(\ref{eq:pol_eq}) can be
 reduced to the quadratic equation of $M^2$.  As a result, we obtain
 trans-magnetosonic solutions easily {\it without the analysis for the
 critical conditions of the magnetosonic points}. \/  
 Although the toroidal magnetic field has several constraints at some
 points along the flow (i.e., the event horizon, the \alfven point, the
 Anchor point, etc), the typical feature of $B_\phi$ for a given $B_p$
 has been discussed in \cite{Takahashi02} by using the standard critical 
 point analysis.  So, we can apply it to make a reasonable test function
 for $B_\phi(r,\theta)$. For example, we can introduce it as a simple
 form of   
\begin{equation}
     \beta^2 = (-g_{\phi\phi})(\Omega_F-\omega_{H})^2 
   \left[\; 1 + C \left( \Delta/\Sigma \right) \; \right] \ , 
\end{equation}   
 where $C$ is a constant. The parameter $C$ is introduced to specify the
 discontinuity of $B_\parallel$ at the shock front, where $B_\parallel$
 is the components of magnetic field parallel to the shock front.

 Now, we will consider shocked MHD accretion onto a black hole. 
 To obtain a shocked black hole accretion solution, it is necessary to
 set up two trans-magnetosonic solutions, which correspond to the
 upstream and downstream solutions.  
 At the shock location, where the shock conditions are required, the
 upstream super-fast magnetosonic solution is connected to the
 downstream sub-fast magnetosonic solution within suitable ranges
 for the field-aligned flow parameters. After the shock, the sub-fast
 magnetosonic inflow passes through the second fast magnetosonic point, 
 and then it falls into the horizon.  In general, the downstream
 solution would have different values for the field-aligned parameters
 from the upstream solution due to the radiation loss at the shock
 front. 
 Hereafter, we assume the conservation of the angular momentum of the
 flow $L$, the angular velocity $\Omega_F$ of the magnetic field line
 and the number flux $\eta$ across the shock front, while the energy 
 of the flow jumps across the shock front; that is 
 $E_{\rm up} > E_{\rm down}$.

 The conditions for the MHD shock formation 
 in the general relativistic framework are summarize by, e.g., 
 \citet{Lichnerowicz67,Lichnerowicz76}; that is, the particle number, 
 the energy momentum and the magnetic flux are conserved across the
 shock \citep[see also][]{AC88,TRFT02,TGFRT06}.  
 For an isothermal MHD shock in cold plasma flow, 
 from the jump condition for the radial component of the energy
 momentum conservation, we obtain the simple relation of 
\begin{equation}
   \left( M^2 + \frac{1}{2} \beta^2 \right)_{\rm up} = 
   \left( M^2 + \frac{1}{2} \beta^2 \right)_{\rm down} \ ,  \label{eq:sk}
\end{equation}
 where for the sake of simplicity we assume that 
 $B_p^{\rm up}=B_p^{\rm down}$, 
 $|B_\phi^{\rm up}| < |B_\phi^{\rm down}|$ (for the fast-magnetosonic
 shock) and the shock front is normal to the upstream flow in the
 poloidal plane.    
 The value for $\beta^2$ (i.e., the value for $C$) increases across the
 fast magnetosonic shock, and then the Mach number decrease; that is,
 the kinetic energy of the flow is released there. 

\bigskip

\begin{figure}[t]
 \begin{center}
   \includegraphics[scale=0.8]{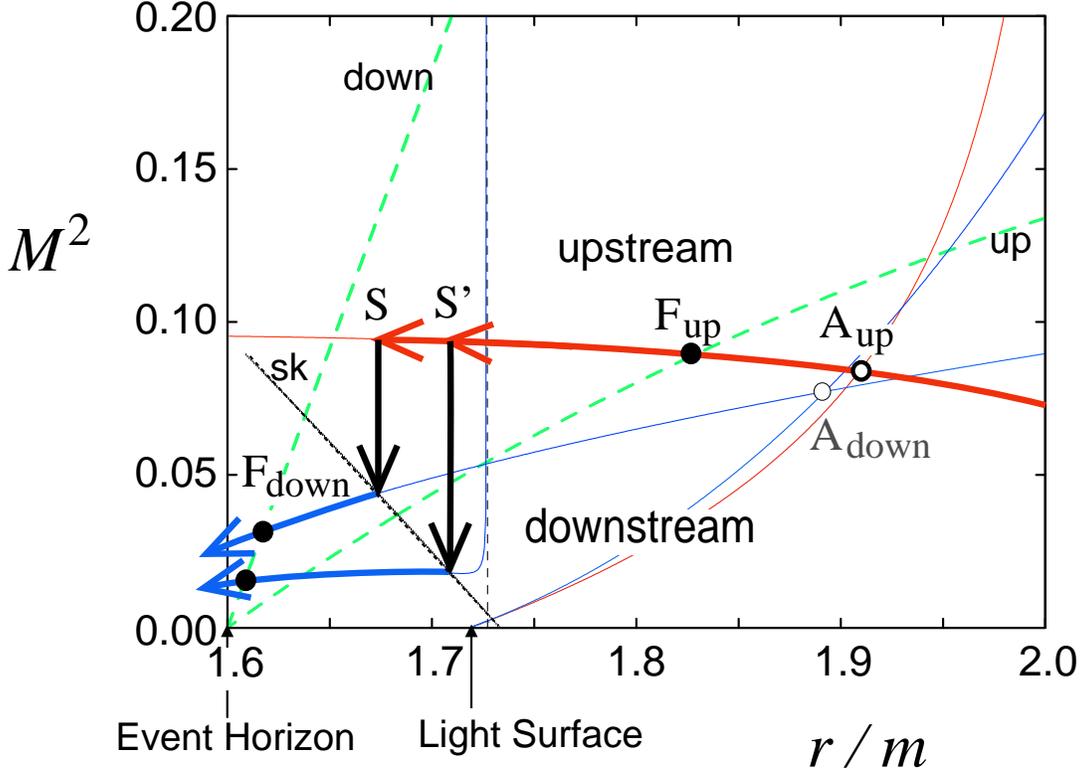}
 \end{center}
 \caption{
 The Alfv\'{e}n Mach number for shocked MHD accretion onto a
 rotating black hole with $a=0.8m$, $\Omega_F=0.4\Omega_{\rm max}$, 
 $\hat{L}\Omega_F=-0.5$ and $\theta=\pi/2$.  
 The preshock solution of $\hat{E}_{\rm up}=1.0$ is plotted by the red
 thick curve, while the postshock solutions for $\hat{E}_{\rm down}=0.8$
 (S) and $\hat{E}_{\rm down}=-0.43$ (S$'$) are shown by the blue
 curves. As a parameter of the magnetic field line, we set 
 $C_{\rm up}=-5.0$ and $C_{\rm down}=40.0$.  
 The broken curves show the \alfven Mach number related to the fast
 magnetosonic wave speed for the upstream and downstream flows. The
 dotted line labeled sk shows the \alfven Much number $M^2_{\rm down}$
 given by eq.(\ref{eq:sk}). The shock radius for each postshock solution
 is shown by the intersection between a blue curve and the dotted black  
 line. 
}
 \label{fig:sk-acc}
\end{figure} 

\begin{figure}[t]
 \begin{center}
   \includegraphics[scale=0.8]{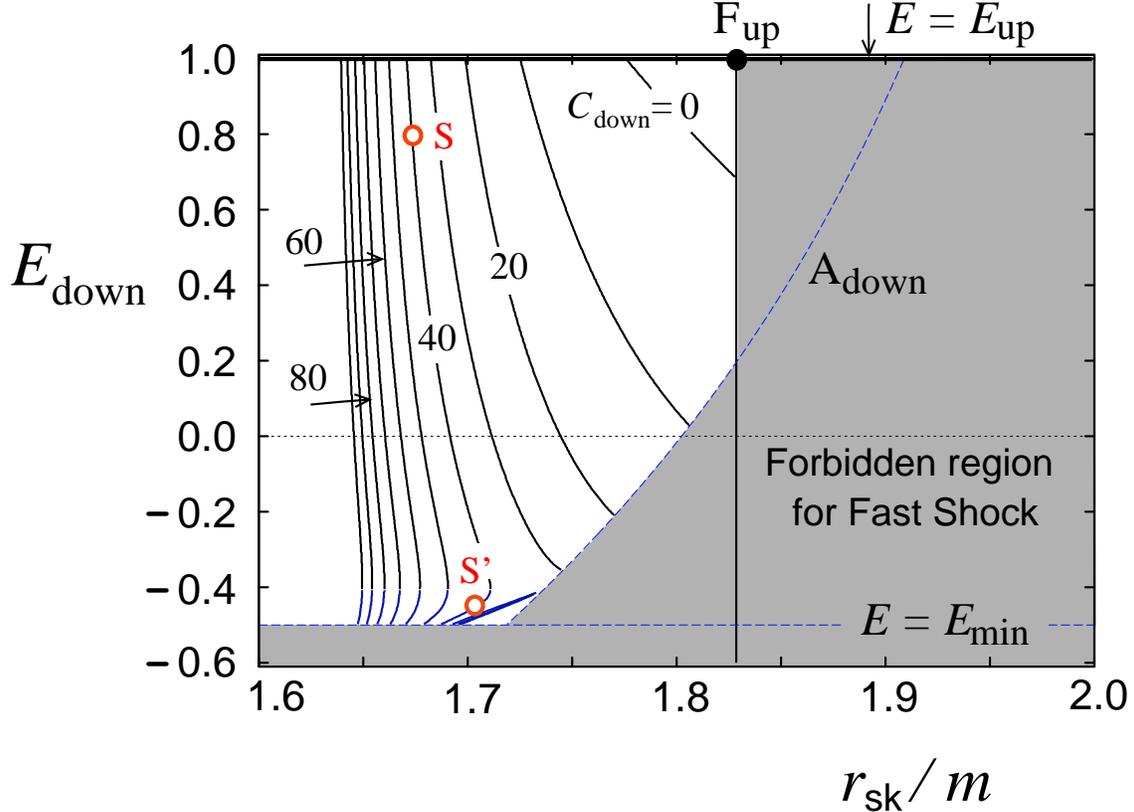}
 \end{center}
 \caption{
 The relation between the energy of the postshock flow and the shock
 location for the solutions for shocked MHD accretion onto a rotating 
 black hole with $a=0.8m$ and $\Omega_F=0.4\Omega_{\rm max}$ (thick
 curves). The values of $C_{\rm down}$ are selected from 0.0 to 100.0,
 while $C_{\rm up}=-5.0$.  
 The shaded region is the forbidden region for the fast MHD shock. 
 The broken curve (A$_{\rm down}$) shows the location of the \alfven
 point for the postshock flow, whose location depends on the 
 energy $\hat{E}_{\rm down}$.  The vertical line shows the location of
 fast magnetosonic point F$_{\rm up}$ of the preshock flow. 
}
 \label{fig:sk-energy}
\end{figure} 

\section{ High Energy Radiation powered by Rotating Black Hole }

 Rotational energy of a rapidly rotating black hole can be extracted by
 MHD inflows with {\it negative}\/ energy due to the interaction between
 magnetic and gravitational fields in the black hole spacetime 
 \citep{BZ77,TNTT90}.  The extracted energy is carried to the
 magnetosphere in the form of outgoing Poynting flux. If the extracted
 energy is converted to some kinds of fluid's energy related to the
 radiative process directly, it may be observable for us. 
 However, in the framework of original Blandford--Znajek (BZ) process
 in the force-free magnetosphere, realistic conversion mechanisms from
 the magnetic energy to the other form are still not clear, because in
 the force-free approximation the inertia of matters is ignored.

 We are now discussing MHD accretion onto a black hole. The MHD inflow
 takes the fluid energy into the hole (that is, the kinetic energy, the
 internal energy and the rest mass energy of the fluid), which is 
 positive at the plasma source.  
 However, the situation of energy extraction from a rotating black hole
 can be also described in the MHD magnetosphere. In this case, the 
 {\it negative energy}\/ MHD inflow ($E<0$) is realized, where it is
 required that the \alfven point of the flow considered is located close
 to the inner light surface in addition to the condition for the BZ
 process; both the \alfven point and the inner light surface need to
 posit inside the ergosphere \citep{TNTT90}.  By considering negative
 energy MHD inflows, the rotational energy of the black hole can be
 carried in the surrounding magnetosphere as the outward energy flux to 
 the distant region directly \citep{MG04,McKinney06} or indirectly by
 way of an equatorial disk \citep{Li02,Wang03}. In the latter case, the 
 extracted energy from the hole is deposited onto the disk, and then it
 converts to the thermal energy of the disk plasma and/or the kinetic
 energy of the outgoing disk winds.

 Now, at the fast MHD shock, we understand that the kinetic energy of
 the upstream MHD flow converts to the thermal and magnetic energies of
 the downstream flow.  The most efficient energy conversion to radiation
 would be achieved by isothermal shock; where it is assumed that the
 thermal energy generated at the shock front escapes from the shocked
 plasma immediately.  
 When the negative energy accretes onto the black hole as the postshock
 flow, we can also expect that the extracted hole's energy would convert
 to the radiative energy at the shock front (see
 Fig.~\ref{fig:BH-Mag}b). 
 By considering the regularity condition at the Alfv\'en point, which 
 is related to the amount of the jump of energy and angular momentum
 between the preshock and postshock solutions, we find the necessary
 condition for such a energy release process 
 \citep[][in preparation]{MT-RT10}.

 Figure~\ref{fig:sk-acc} shows two examples of solutions for shocked
 accretion onto a black hole.  After passing through the \alfven point  
 A$_{\rm up}$ and fast magnetosonic point F$_{\rm up}$, the accretion
 flow becomes super-fast magnetosonic, and then it falls into the black
 hole.  When the MHD shock arises at the radius labeled S or S$'$ on
 the accreting flow (see the vertical downward-arrows), the postshock
 flow becomes sub-fast magnetosonic. Then, the postshock flow must
 become super-fast magnetosonic again by passing through the fast
 magnetosonic point F$_{\rm down}$ or F$_{\rm down}'$ located between
 the shock front and the event horizon. 
 Note that the branch of the postshock solution labeled S connects to 
 the \alfven point A$_{\rm down}$, and it is also a shock-free accretion   
 solution.  On the other hand, the postshock flow solution labeled S$'$
 shows an unphysical solution as a shock-free solution, because the 
 branch of the solution does not connect to the \alfven point and
 diverges outside the shock location.  We can not accept such a
 divergent solution as a shock-free solution streaming from the plasma
 source to the horizon.  However, by considering the shock formation,
 this solution S$'$ can survive as a physical postshock flow solution in
 the region within the shock radius.

 At last, we consider the conditions for the formation of the postshock
 solution with negative energy and the possible range of the amount of
 the energy extracted from the black hole. 
 Figure~\ref{fig:sk-energy} shows the relation between the energy of 
 the postshock flow $\hat{E}_{\rm down}$ and the shock radius 
 $r_{\rm sk}$ for the accretion flow onto the black hole with 
 $a=0.8m$.  The energy of the preshock flow solution is set  to
 $\hat{E}_{\rm up}=1.0$ and the angular velocity of the magnetic field
 lies is set to $\Omega_F = 0.4 \Omega_{\rm max}$, where 
 $\Omega_{\rm max}$ is the maximum value for the field line existing 
 the plasma source region located between the inner and outer light
 surfaces \citep[see][]{TNTT90}.  Some thick curves in
 Figure~\ref{fig:sk-energy} correspond to the different values of 
 $C_{\rm down}$ for the postshock solution, 
 and we set $C_{\rm up}=-5.0$ for the preshock flow solution.  
 Although we consider the case where the shock front is generated inside
 the ergosphere, the postshock flow with negative energy ($E<0$) is 
 possible. This means the extraction of the hole's rotational energy by
 the ingoing MHD flow.  The minimum energy of the postshock flow 
 $E_{\rm min}$ is given by $E = L\Omega_F < 0$. 
 The extracted energy can be released at the shock front where the hot
 plasma region is generated. From the hot plasma in the shock front, the 
 observable energy output is expected; e.g., in the form of high-energy
 radiation, bulk-motion of outgoing plasma flows, etc.

\section{ Concluding Remarks }

 We have discussed global shocked accretion solutions in a black hole
 magnetosphere, which are composed of two trans-magnetosonic solutions
 with a MHD shock where the general relativistic MHD has been applied to
 the accretion flows.  Then, we have shown the possibility of a very hot 
 shocked plasma region very close to the event horizon. 
 At this stage, in the study of the black hole magnetosphere, the
 expected energy spectrum from the MHD shock is not clear.  
 Nevertheless, we newly point out the possibility of the energy
 extraction from the black hole through the MHD shock with the negative 
 energy in the postshock flow. 
 Although, in this paper, we show the possibility of shocked accretion
 solutions with negative energy ($E<0$), the dependence of flow   
 parameters (i.e., $E_{\rm up}$, $\Omega_F$, $a$, $L_{\rm up}$) and
 flow's latitude will be presented in the next paper (Takahashi \&
 Takahashi, in preparation).  

 To complete the black hole's rotational energy release at the MHD
 shock, we must investigate the structure of the shocked hot plasma
 region.  Although the value of the released energy depends on the
 field-aligned flow parameters, we can estimate the plasma density and
 temperature there by the help of the accretion disk models that give
 the boundary condition for the magnetosphere.  We need a realistic
 model of the black hole magnetosphere with a magnetized accretion disk.  
 Although we assume a given magnetic field line, the configuration may
 be unstable. Then, the dynamical behaviors of the field configuration
 should be also investigated by helping numerical simulations
 \citep[e.g.,][]{Koide06,MG04,McKinney06}. Then, 
 a significant amount of X-rays and gamma-rays would be estimated. 
 In order to take into account such radiation effects in the general
 relativistic plasma, we should formulate the local radiation effects of
 MHD fluid by using the energy-momentum tensor with the local heat flux, 
 in addition to the fluid part and the electromagnetic part.  
 Then, we will find the evidences for {\it real}\/ black holes 
 in the observational data 
 (e.g., the spectrum in high-energy range, the short time variability,
 the radio image of black hole shadow, X-ray polarization, etc). 
 That is, we expect that the high energy emission from this hot plasma
 brings to us {\it additional}\/ information about the black hole
 spacetime; especially, the polar region emission including this
 information can be distinct from the emissions from the equatorial
 plasma source, which have been investigated by many authors in the models 
 of accreting gas disk.


\acknowledgments

 We are grateful to Sachiko Tsuruta for her helpful comments.  
 This work was supported in part by the Grants-in-Aid of the Ministry 
 of Education, Culture, Sports, Science and Technology of Japan
 [19540282 (M.T.), 21740149 (R.T.)].


Facilities: ***



\end{document}